\documentclass[aps,prb,floats]{revtex4}

  \font\elevenmib=cmmib10 scaled 1095
  \font\tenmib=cmmib10
  \font\eightmib=cmmib10 scaled 800
  \font\sixmib=cmmib10 scaled 667
  \skewchar\elevenmib='177
  \newfam\mibfam
  
  \textfont\mibfam=\tenmib
  \scriptfont\mibfam=\eightmib
  \scriptscriptfont\mibfam=\sixmib
\normalsize
\usepackage{ifthen}
\usepackage{ifpdf}

\usepackage{latexsym}
\usepackage{amsmath}
\usepackage{amssymb}
\usepackage{bm}
\usepackage{wasysym}
\usepackage{amsfonts}
\usepackage{graphicx}
\usepackage{subfigure}
\usepackage{amssymb}
\usepackage{bm}
\usepackage[dvips]{color}
\usepackage{longtable}
\usepackage{hyperref}
\ifpdf
\usepackage{epstopdf}
\else
\usepackage{epsfig}
\fi

\newcommand{\hide}[1]{}

\begin{document}
\title{Broken Time-reversal Symmetry in Josephson Junction with an Anderson impurity and multi band superconductors }
\author{Y. Avishai$^{1,2}$ and T.K. Ng$^1$}
\email {yshai@bgu.ac.il}
\email {phtai@ust.hk}
\affiliation{$^1$ Department of Physics, Hong Kong University of Science and Technology, Clear Water Bay, Kowloon, Hong Kong\\
$^2$ Department of Physics and Ilse Katz Institute for Nanotechnology, Ben Gurion University of the Negev, Beer Sheva, Israel}

 \begin{abstract}
 A Josephson junction consisting of an Anderson impurity weakly coupled with two-band and  single-band superconductors exposes a time reversal breaking ground state when  the coupling between the two bands  exceeds a certain threshold. The critical regime
occurs around local moment formation.
 This indicates a fundamental and distinct role of strong correlations: Driving a system into a time reversal breaking ground state. One of the observable consequences is that the impurity magnetization in this phase is reduced. 
  
\end{abstract}
\maketitle
\narrowtext
\noindent
{\bf Motivation}: Recently, there is a renewed interest in the physics of multi-band superconductors which are believed to encode the
 physics behind  the new Iron based superconductors \cite{Raghu,Lee,TK}.
 Historically, interest in two-band superconductors is related to experiments suggesting that in some transition metals,
 there should be a second mechanism responsible for pairing beyond that of electron-phonon
 interaction\cite{Suhl, Muskalenko,Chaw, Kondo}.  For example, in Nb$_3$Sn the isotopic effect is found to be very
 weak\cite{Devlin}. Inter-band coupling also provides a natural mechanism for augmenting $T_c$.\cite{Kondo,Ora}
 
 \noindent
 The physics associated with a two-band superconductor can be examined
 in a Josephson junction involving two-band superconductors on one of its sides and a third
 superconductor on the other side \cite{TKN}.  Within a Ginzburg-Landau formalism
the order parameters $\psi_1, \psi_2$ of the two
 bands are coupled among themselves by a term $-J$Re$[\psi_1 \psi_2^*]$, and to the
 single-band superconductor (on the opposite side of the junction) with tunneling strengths $\Gamma_1$ and $\Gamma_2$.
 A time-reversal violating ground state (TRVGS) was found for $\Gamma_1 \approx \Gamma_2$,
 due to frustration between the three order parameters, 
 while otherwise, a time reversal conserving ground state (TRCGS) prevails.
A natural question which motivates the present research is what would happen if the Josephson junction contains an Anderson
 impurity.  Indeed, strong correlations are expected to affect these findings and alter the pertinent physics in a profound way. 

\begin{figure}[!h]
 \centering
 \includegraphics[width=7.4truecm]{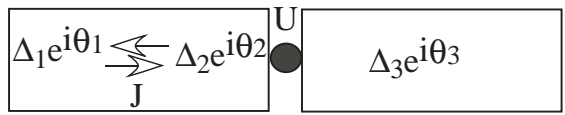}
 \caption{Geometry of the $2B-U-1B$  junction}
 \label{fig1}
 \end{figure}

 \noindent
 {\bf In this work} we study a system consisting of an Anderson impurity 
 (level energy $\varepsilon$ and correlation strength $U$) weakly coupled on the left with a two-band
 superconductor (order parameters $\psi_n=\Delta_n e^{i \theta_n} ,n=1,2$ and inter-band Josepshon coupling $J$)
 and on the right to a single-band superconductor (order parameter $\psi_3=\Delta_3 e^{i \theta_2}$).
  This is briefly denoted as $2B-U-1B$ Josephson junction (see Fig~\ref{fig1}).
 A TVRGS occurs if the free energy ${\cal F}[{\bm \theta}\equiv (\theta_1, \theta_2, \theta_3)]$ has a minimum at some
 point ${\bar {\bm \theta}}$ for which there is a non-zero Josephson current loop ${\cal J}_c$ even in the absence of
 magnetic field (see  Ref.~\cite{TKN} for the case of  $2B-1B$ Josephson junction without Anderson impurity).
 On the technical part, the mean-field (Hartree-Fock) approach to the $1B-U-1B$ problem \cite{Rozhkov,AG} can be used
 to treat the $2B-U-1B$ system as well.  It enables the elucidation of the ground-state configuration ${\bar {\bm \theta}}(J,U)$
 together with the free energy ${\cal F}({\bar {\bm \theta}})$, the Josephson current ${\cal J}_c({\bar {\bm \theta}})$
the impurity occupation $n_{{\mathrm imp}}({\bar {\bm \theta}})$ and magnetization $m_{{\mathrm imp}}({\bar {\bm \theta}})$. As it turn out, however, the physics is distinct since there are now three order parameters 
and two of them are coupled.

\noindent {\bf Our main result} pertains to the nature of the ground state TR symmetry in the $U-J$ plane,  for high asymmetry parameter,  $\delta \equiv\frac{\Gamma_1}{\Gamma_2} \gg 1$,
 where no TRVGS exists\cite{TKN} on the line $U=0$.
Two phases are identified, one for which there is TRCGS and
 one for which there is TRVGS, separated by a sharp border line. The TRVGS occurs for
 $U_1(\delta)<U<U_2(\delta)$ and for $J>J_c(U, \delta)>0$. Here $U_1<U_0$ and $U_2>U_0$ where
$U_0$ defined by  $\varepsilon+\frac{U_0}{2}=0$ is the particle hole symmetric point
and $J_c(U, \delta)$ is some critical
 threshold. \\
 {\bf Formalism}: The mean-field Hamiltonian is written as,
 \begin{equation} \label{Hamiltonian}
H=H_{\mathrm{L}}+H_{R}+H_{\mathrm{I}}+H_{\mathrm{imp}}.
\end{equation}
The first two terms $H_L, H_R$ describe (within
the BCS formalism) the two band superconductor on the left ($H_L$)
and the single band superconductor on the right ($H_R$). Explicitly, in terms
of quasi-particle field operators,
$H_{\mathrm{L}}=\sum_{n=1}^{2} H_{n}$ and $H_{R}=H_3$ are written as,
\begin{equation} \label{Hn}
H_{n}=\int \Psi_{n}^{\dagger}({\bf r}) {\cal H}_{n} \Psi_{n}({\bf r}) d{\bf r}, \quad~ \Psi_{n}({\bf r})=\begin{bmatrix} \psi_{n \uparrow}({\bf r}) \\ \psi_{n \downarrow}^{\dagger}({\bf r}) \end{bmatrix},
\end{equation}
The two-band Hamiltonian densities ${\cal H}_{n=1,2}$ are derived rom the two-band model Hamiltonian of Ref.~\cite{Kondo}.  In this Hamiltonian (equation
(1) therein), there is only intraband pairing (and no interband
pairing) and the electrons in the two bands are coupled only
through an interband Josephson effect (see below). The single band Hamiltonian density ${\cal H}_3$ has the standard Bogoliubov-De Gennes structure. Thus,
\begin{eqnarray}  \label{H2B}
&& {\cal H}_{n}=\begin{pmatrix} \varepsilon_n(-i {\bm \nabla})-\mu & \Delta_n e^{i \theta_n}+J_{m}e^{i \theta_{m}} \\ \Delta_n e^{-i \theta_n} +J_{m}e^{-i \theta_{m}}& -\varepsilon_n(-i {\bm \nabla})+\mu \end{pmatrix},  \ \
 {\cal H}_3=\begin{pmatrix} \varepsilon_3(-i {\bm \nabla})-\mu & \Delta_3 e^{i \theta_3} \\ \Delta_3 e^{-i \theta_3} & -\varepsilon_3(-i {\bm \nabla})+\mu \end{pmatrix}.
\end{eqnarray}
In Eqs.~(\ref{H2B}) $m \ne n=1,2$ and the various quantities are
defined as follows: $\varepsilon (-i {\bm \nabla})$ is the kinetic
energy operator derived from the corresonding energy dispersion
functions $\varepsilon_n({\bf k})$, and $\mu$ is the chemical
potential. Moreover, $\Delta_n e^{i \theta_n} \equiv V_n \psi_n$
where $\psi_n \equiv \int d{\bf r}  \langle \psi_{n
\downarrow}({\bf r})  \psi_{n \uparrow}({\bf r}) \rangle$ is the
order parameter of superconductor $n=1,2,3$ and $V_n$ is the
corresponding strength of the pairing potential. Similarly,
$J_{m}e^{i \theta_m} \equiv I \psi_{m}$ encodes the pairing field
in band $n \ne m$ due to electrons pairing in band $m$ (interband
Josephson effect) and $I$ is the strength of the coupling between
the two bands \cite{Kondo}.

 \noindent The tunneling part $H_I$ contains hopping between the impurity to each one of the three superconductors (with
 different strengths $t_{n=1,2,3}$), which occurs at a single point. Finally, the strong correlation part has the usual structure of an Anderson impurity Hamiltonian.
Explicitly,
\begin{equation} \label{HI}
 H_{\mathrm{I}}=-\sum_{n=1}^{3}t_n[\Psi_{n}^{\dagger}({\bf 0})\tau_3 C+C^{\dagger} \tau_3 \Psi_{n}^({\bf 0})+h.c], \quad~ C=\begin{bmatrix} c_{\uparrow} \\ c^{\dagger}_{\downarrow} \end{bmatrix}, \ \
H_{\mathrm{imp}}={\bar \varepsilon} C^{\dagger} \tau_3 C+\frac{1}{2}U [C^{\dagger}C]^{2}, \quad~ {\bar \varepsilon}=\varepsilon+\frac{U}{2}.
\end{equation}

The procedure for calculating the free energy ${\cal
F}(\theta_1,\theta_2,\theta_3)$ is a modified version of the
algorithm used in Refs.~\cite{Rozhkov,AG}. It involves an
Eucledean path integral for the partition function $Z=e^{-\beta {\cal F}}$
 in terms of
Grassman fields and employing Hubbard-Stratonovich transformation
for treating the quartic term in $H_{\mathrm{imp}}$ at the expense
of an additional integration on a new field $\gamma$. The latter
is carried out within the saddle point approximation leading to a
self consistent equation (Hartree-Fock approximation). Notice that
unlike Ref.~\cite{TKN} where the pairing field away from the
impurity is also solved self-consistently within a Ginzburg-Landau
approximation, the pairing fields $\Delta_n$'s are assumed to take
a constant value in the superconductors.

Beside the density of states at the Fermi energy $N(\mu)$ (assumed
constant) and the impurity level (partial) widths $\Gamma_n=\pi
t_n^2 N(\mu)$ for the superconductor $n=1,2,3$, the basic input
quantities are defined below, [with
$\omega=\omega_k=(2 k+1) \pi T$ ($k=0,1,2, \dots$) a Matsubara
frequency at temperature $T=1/\beta$  and $\sum_\omega f(\omega)
\equiv \sum_k f(\omega_k)$]:
\begin{eqnarray}
&& \alpha_{n=1,2} (\omega) = \Gamma_n[\Delta_n^2+J_{m}^2+2 \Delta_n J_{m} \cos(\theta_1-\theta_2)+\omega^2]^{-\frac{1}{2}}, \ \
\alpha_{3} = \Gamma_3[\Delta_3^2+\omega^2]^{-\frac{1}{2}}, \ \  \eta (\omega) =\omega [1+\sum_{n=1}^3 \alpha_{n}(\omega)]. \nonumber \\
&& q_{1}(\omega)=\alpha_{1}(\omega) \Delta_1+\alpha_{2}(\omega) J_1, \ \
q_{2}(\omega)=\alpha_{2}(\omega) \Delta_2+\alpha_{1}(\omega) J_2, \quad~ q_{3}(\omega)=\alpha_{3}(\omega) \Delta_3. \nonumber  \\
&& F(\omega; {\bm \theta}) \equiv \sum_{n=1}^3 q_{n}^2
 +2 \sum_{n \ne n'} q_{n}q_{n'} \cos(\theta_n-\theta_n').
 \label{FF}
\end{eqnarray}
The self consistent equation for the  field $\gamma$ (analogous to Eq.~7 in Ref.~\cite{Rozhkov} or Eq.~12 in Ref.~\cite{AG}) reads,
\begin{equation} \label{saddleT0}
  \frac{1}{2U}-T \sum_\omega \frac{[\gamma^2+\eta(\omega)^2-{\bar \varepsilon}^2- F(\omega, {\bm \theta})]}{ [\gamma^2-\eta(\omega)^2-{\bar \varepsilon}^2- F(\omega,{\bm \theta})]^2+4 \gamma^2 \eta(\omega)^2}=0,
  \end{equation}
  whose solution  ${\bar \gamma}({\bm \theta})$ is used below.

 \noindent
{\bf Analysis of the results}: The free energy associated with the impurity is the coefficient of $-\beta$ in the exponent of
 the partition function $Z=e^{-\beta {\cal F}}$, and the formalism described above yields for it  the following expression,
\begin{eqnarray}
&& {\cal F}({\bm \theta})=\frac{{\bar \gamma}^2}{2 U} + {\bar \varepsilon}
 -\frac{1}{2}T\sum_{\omega}  \ln \beta^4 \{ [{\bar \gamma}^2-\eta(\omega)^2-{\bar \varepsilon}^2- F(\omega,{\bm \theta})]^2+4 {\bar \gamma}^2 \eta(\omega)^2 \}.
 \label{GS}
\end{eqnarray}
It will be examined as function of  $J$ and $U$ fixing other parameters as,
  \begin{equation} \label{parameters}
  \Delta_{1,3}=1.0, \ \Delta_2=0.8,  \ \Gamma_1=0.5, \ \Gamma_2=\Gamma_3=0.2, \ \varepsilon=-2, \ \ T=0.0005,
 \end{equation}
 (energies are expressed in unit of $\Delta_1=1$).
Note the ratio $\delta \equiv \frac{\Gamma_1}{\Gamma_2} =2.5$ and recall that for noticeable different Josephson tunneling strengths $\Gamma_1 \ne \Gamma_2$,
 there is no TRVGS in a $2B-1B$ junction\cite{TKN} .

\noindent
 Employing gauge invariance and setting
 $\theta_3=0$, the minimum ${\bar {\bm \theta}}=({\bar \theta_1},{\bar \theta_2}, 0)$ of  ${\cal F}( {\bm \theta})$ is
 located.  If $\sin {\bar \theta}_1=\sin {\bar \theta}_2=0$ one evidently  has a TRCGS where
 the Josephson current vanishes. On the other hand, when the above condition is not satisfied, one has a TRVGS. In the
 {\em absence of Anderson impurity} and close to perfect symmetry $\delta \approx 1$ it occurs for $J>0$ at ${\bar \theta}_1 \approx-{\bar \theta}_2 \approx \frac{\pi}{2}$.
 For
 $U \ne 0$ the property  ${\bar\theta}_1 \approx -{\bar \theta}_2 \approx \frac{\pi}{2}$ is expected to be somewhat
 modified, and regions with $\sin {\bar\theta}_1,\sin {\bar \theta}_2 \ne 0$
 are those where TRVGS is realized.  
 
\begin{figure}[!h]
\centering
\includegraphics[width=5.5truecm,angle=90]{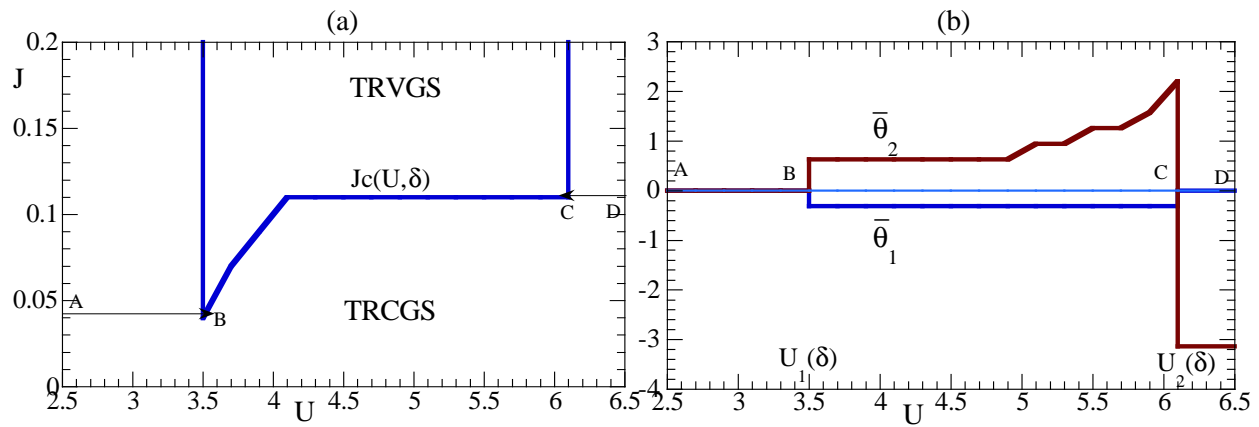}
\caption{
 (a) Phase diagram showing the line $J_c(U, \delta)$ separating TRCGS from TRVGS in the $U-J$ plane.
 All parameters (except $U$ which varies) are defined in Eq.~\ref{parameters}. Note the sharp cutoff points at $U_1(\delta) \approx 3.5$ (below the point $B$),
 and $U_2(\delta) \approx 6.1$ (below the point $C$.) (b) The phases ${\bar \theta}_1, {\bar \theta}_2$ of the order parameters at the ground state configuration
(minimum point of the free energy, Eq.~(\ref{GS})), are displayed along
the line $ABCD$ of Fig.~\ref{fig2}a. Along the border line $J_c(U, \delta)$ we have $\sin {\bar \theta}_1, \sin {\bar \theta}_1 \ne 0$.
}
\label{fig2}
\end{figure}

 \noindent
 The phase diagram displaying these two domains is shown in Fig.~\ref{fig2}a.
The order parameter phases of the TRVGS just above the border line $J_c(U, \delta)$ are displayed in Fig.~\ref{fig2}b, and indeed, they are neither $0$ nor $\pi$ so that $\sin {\bar \theta}_1, \sin {\bar \theta}_1 \ne 0$.
The scenario emerging from Fig.~\ref{fig2} is as follows: For a given asymmetry parameter
 $\delta \gg 1$ there is a domain in the $U-J$ plane for which TRVGS is realized. It is 
 bounded on the left and right by vertical lines $U=U_1(\delta)$ and $U=U_2(\delta)$ and below by a border line $J_c(U, \delta)>0$. It is checked that
 $U_1(\delta) <U_0=2 |\varepsilon|$ is a slowly increasing function of $\delta$ starting at $U_1(1)=0$ while $U_2(\delta)>U_0$ is very weakly dependent on $\delta$.  Moreover,   $J_c(U,\delta)$ is an increasing function
 of $U$ and $\delta$ (actually it saturates around $U_0$).

\noindent
  To understand the mechanism which drives the formation of the TRVGS state, we first consider the case of single-band superconductor
  Josephson junction\cite{Rozhkov}. It was found that the effective Josephson
  coupling between the two superconductors mediated through the Anderson Impurity changes sign when $U$ goes through a
  critical value $U_c$. The two superconductors are in phase ($\delta=0$) at $U<U_c$, and are out of phase ($\delta=\pi$) at
  $U>U_c$. Associated with this transition is the formation of a large magnetic moment at the Anderson impurity at
  $U>U_c$. We now replace one of the superconductor by a two-band superconductor. In this case the effective Josephson
  coupling $F_{\mathrm{Jos}}$ between the different $\theta_n$'s generated by the impurity has the form
  \[
    F_{\mathrm{Jos}}\sim
    T_{13}(U)\cos(\theta_1-\theta_3)+T_{23}(U)\cos(\theta_2-\theta_3)+T_{12}(U)\cos(\theta_1-\theta_2)+
    O(\cos(\theta_n)\cos(\theta_m))+...,
    \]
   where $T_{13}(U)$ and $T_{23}(U)$ become small around $U\sim U_c$ and higher order terms in $F_{\mathrm{Jos}}$ becomes important in
   determining the phase structure. The higher order Josephson terms arises from $\alpha_n(\omega)$'s and
   $F(\omega; {\bm \theta})$ in Eq.\ (\ref{FF}) which has much more complicated structure than the case of one-band
   superconductor considered in Ref.\cite{Rozhkov}.  Apparently, the coupling between the superconductors and the formation of $\pi$ junction for large
   $U$ introduces additional frustration among the phases $\theta_n$ which gives rise to TRVGS when the first order coupling
   terms $T_{13},T_{23}$ become small, i.e. around the critical regime $U\sim U_0$.

\begin{figure}[!h]
\centering
\includegraphics[width=5.5truecm,angle=90]{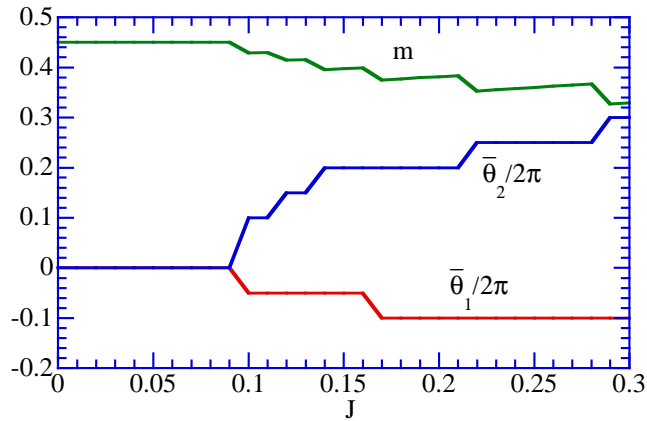}
\caption{
 Impurity magnetization $m$ (upper line) at the ground state is displayed as function of $J$ for $U=4$ with other parameters defined in Eq.~(\ref{parameters}). Also displayed are the phases $({\bar \theta}_1,{\bar \theta}_2)$ (divided by $2 \pi$) of the order parameters indicating the ground state configuration.
Both the magnetization and the phases undergo a remarkable change of behavior
 at the border line where $J=J_c(U,\delta)$ (according to Fig.~\ref{fig2}a $J_c(U=4,\delta) \approx 0.1$). The magnetization becomes smaller due to frustration in the TRVGS phase and the commensurability of steps indicates that the system 
 passes through higher and higher frustrated ground-states. At the same time, 
 it is verified that the ground state energy is a very smooth function of both $J$ and $U$.
}
\label{fig3}
\end{figure}

  \noindent
  The question now is whether this phase where TRVGS occurs can be traced experimentally.
  A promising direction is to inspect the impurity magnetization $m$ and its behavior as the phase boundary is crossed.
  Specifically, we fix the repulsive potential $U$ and inspect the magnetization as
  $J$ is varied, thereby crossing the phase boundary at $J_c(U,\delta)$. In Ref.~\cite{Rozhkov} it was found that for a
  1B-U-1B junction, the ground state energy and the magnetization develop kinks near the critical point (when studied as
  function of the phase difference between left and right superconductors). Here, on the other hand, the behavior of the
  ground state energy (as function of $J$) appears to be very smooth and rather slowly varying. It should be stressed that
  here we do not vary the phases of the order parameters, 
  as they are fixed at the ground state position $({\bar \theta}_1,
  {\bar \theta}_2)$. The mean field magnetization
  $m=\frac{2 {\bar \gamma}}{U}$ and the ground state order-parameter phases $({\bar \theta}_1,{\bar \theta}_2)$ are displayed as function of $J$ for $U=4$ in Fig.~\ref{fig3}.
The magnetization in the TRVGS phase is smaller than in the TRCGS phase because
according to the spin analog developed in Re.~\cite{TKN} the former phase
is characterized by frustration.  The fact {that ${\bar \theta}_{1,2}$ also
  shows step behavior as $J$, and that  these steps are commensurate with those of the magnetization  lead us to conjecture that the system has many frustrated states close in energy.

\noindent
{\bf Conclusions:}
 The study of $2B-U-1B$ carried out here is motivated by the renewed interest in multi-band superconductors (stemming from the analysis of  the Iron based superconductors).
 The pertinent physics is fundamentally distinct from that of a $1B-U-1B$ system\cite{Rozhkov,AG}
 and the $2B-1B$ junction discussed in Ref.\cite{TKN} since the roles of strong correlations $U$ and
 the coupling $J$ between the two order parameters in the two-band superconductor interlace.
 It has been shown that for $J > J_c(U, \delta)>0$ and for $U_1<U<U_2$ a TRVGS emerges which
 supports a non-zero Josephson current even without a magnetic field. The role of strong correlations as
 controlling the TR symmetry of the ground-state is evidently remarkable.
 As far as an experimental detection is concerned, beside the experiments proposed in Ref.\cite{TKN}, our preliminary
 results indicate that
 some quantities (such as impurity magnetization) undergo a dramatic change as $J$ passe through the phase boundary $J_c(U,\delta)$, hence
 a Josephson junction with Anderson impurity can serve as a potential tool for probing the relative phase of the two order
 parameters in a two-band superconductor (which is a very elusive quantity).

\noindent
{\bf Acknowledgement:} We would like to thank A. Golub, O. Entin-Wohlman and F. C. Zhang for fruitful discussions.  The work of Y.A is partially supported by an ISF grant.


\end{document}